\documentclass[twocolumn,aps,nofootinbib]{revtex4-1}
\usepackage{graphicx}
\usepackage{booktabs}
\usepackage{xcolor}
\usepackage{hyperref}
\usepackage{multirow}
\usepackage{setspace}
\usepackage{amsmath, amssymb, amsthm}
\usepackage{cleveref}
\usepackage{physics}
\usepackage{url}
\newcommand{\comment}[1]{}

\begin{document}
\date{\today}
\author{Sunil Kumar Raina MD FIAPSM$^1$, Yaneer Bar-Yam$^2$} 
\affiliation{$^1$Dr RP Government Medical College, Tanda (HP)}
\affiliation{$^2$New England Complex Systems Institute, Cambridge, MA USA}
\title{Was India saved by staying below the critical travel threshold and was lockdown and travel restriction the most important public health intervention?}

\begin{abstract}
Indian response to the pandemic has been described from ``India is in denial about the covid-19 crisis'' or ``India staring at corona virus disaster'', to ``The mystery of India's plummeting covid-19 cases''. These responses have been far from being backed scientifically and appear ignorant of India's capabilities of leveraging its strengths to mitigate the impact of the pandemic. Backed by a swift Government action of restricting/regulating movement to increasing public health capacity to meet the increasing demands of the pandemic, India seems to have done enough to emerge successful. India is doing well, if not guaranteed for the future, but at least for now. Here we review these measures and point to their consistency with analysis of the role of intercommunity transmission and within community action to stop localized outbreaks. In particular, severe restrictions on travel, stopping gatherings, targeted localized lockdowns, school closures, effective public communication, improvements in case identification, rapid ramping of industrial production of masks and other personal protective equipment (PPE) and testing capacity, as well as intensive measures in high density areas of urban deprivation have placed India in a regime of declining cases and outbreak control. It is time to recognize the scientific basis of India's success and give it its due. With the number of new cases in India leveling recently, the urgency is great to complete the eliminiation process so that a new surge does not occur. 
\end{abstract}


\maketitle

\section{Background}

Over one year into the COVID-19 pandemic there continue to be widely speculative ill-informed discussions of causal factors in pandemic dynamics despite scientific clarity about the essential role of public health measures. Some of the most common questions doing the rounds have been; was use of stringent lock down measures necessary? Why did it strike different countries differently or different states/cities in the same country with varying severity? In between such questions there have been headlines that read that ``India is in denial about the COVID-19 crisis'' or ``India staring at corona virus disaster''. [1] Surprisingly however, the last few months have seen a new narrative on India. The headlines now read ``The mystery of India's plummeting covid-19 cases'' although the word ``mystery'' is far from flattering.[2] From India's efforts in mitigating the impact of Covid-19 pandemic being viewed with suspicion to India's success now being equated with magic, the cycle of mis-information on India seems complete. Epidemiological experts and the mathematical modelers alike were predicting millions of deaths by August 2020, the falling numbers have left them perplexed and seeking non-scientific or still speculative explanations such as a hygiene/microbiome theory that has been widely talked about as the factor working in India's favor [5] and the country's age profile having played a positive role---which perhaps contributed to a lower death rate but does not explain why cases have declined. The achievement has been spectacular. As on 8th of February, India, with a population of $1.4$ billion, has cumulative $155,080$ COVID-19 deaths with 84 deaths in the preceding 24 hours and total number of active cases is at $148,609$ is doing much better than in comparison to, the United States, which has seen $439,830$ deaths or the United Kingdom for example, countries boasting of a superior healthcare delivery system. [3] We note that no positive correlation between superior healthcare delivery systems with pandemic excellence has been shown, and indeed, there is no reason to attribute effectiveness of prevention through stopping transmission to the medical quality or capacity of hospitals for treatment of infected individuals. Public health prevention and medical treatment are entirely different capabilities.

\section{The evidence so far}

India recorded 97,859 on 16th September 2020, the highest single-day number of Covid-19 infections since the start of the pandemic. However, ever since then the cases have seen a significant decline despite the number of tests being carried out continuing to be substantial. The Indian Council of Medical Research (ICMR) backed testing strategy had yielded 8 crore (tens of millions of) tests till 30th September out of which 3.1 crore were done in September only with a daily average of $10,42,750$ tests. As recent as 13th of February, the total number of tests conducted was $6,97,114$ to add to a cumulative total of $20,62,30,512$. The number of tests returning positive for Covid-19 was 11,106. The average number of new cases for last one week has been about $11,000$/day. [4]

The number of active cases has also seen a continues decline since 17th September 2020. India recorded highest number of daily deaths on 15th September at $1283$ (except $2006$ on 16th June; which was due to adjustment of old unrecorded deaths) and since then has seen a significant decline. [4]

\section{What has worked in India's favor} 

In an attempt to look into what may have worked for India, we list a few important underlying factors probably favorable to mitigating the impact of the pandemic.

\begin{enumerate}
\item Experience and successful response efforts in prior pandemics has been in India's favor. India saw lesser number of cases of 2009 H1N1 pandemic due to H1N1pdm09 virus when compared to the United States or the Europe or even Africa for that matter. [6] It even saw lesser cases than China, Japan or South Korea.[6] The evidence of history also reveals India's ability to largely restrict pandemics/epidemics to one or few geographical areas only. Even the Spanish flu of 1918 which is supposed to have hit India badly killing an estimated 10 million to 20 million---roughly $6\%$ of the Indian population at that time---did not present a different picture, with the then state of Bombay affected badly in comparison to Bihar for example. [7] Similar observations appear to be the story for seasonal Influenza A (H1N1) where the cases are differentially distributed across the country.[8] We note that Covid-19, did affect India widely and registered a large number of cases. However, here again the highest contribution to cases in this pandemic has also been restricted to some regions only. The reasons for the observed geographical restriction can be attributed to both underlying social conditions as well as proactive implementation of public health measures as enumerated below.
\item The less urbanized India may have worked to its advantage. Of all the continents in the world, Asia is the least urbanized, with $51\%$ of its population living in the urban areas. [9] When it comes to India, the percentage falls further with only around $30\%$ of Indians living in Urban areas. [10] When compared to Europe or America, the difference is significant. A large portion of Indian population survives on agriculture as main source of income generation. This results in limited travel and thus limited rates of community-to-community transmission and the opportunity to implement localized actions to stop outbreaks from spreading further.
\item Another important factor that probably worked in India's favor is the way Indians travel. India continues to use buses as the major public transport mode of travel. Rapid transit (metro/subways/undergrounds etc) travel mode is restricted to a few major cities of the country only. So is travel by the airways. Added to it is the fact that large India states like the Uttar Pradesh or Bihar which account for almost 1/4th of India's population travel less in comparison to states like Kerala or Himachal Pradesh leading to less overall mean travel time by Indian population in general. As per a report, $86\%$ of Indian households never take to a trip.[11] The limited modes of travel both limit transmission between communities and allow for its regulation when proactive measures are taken.
\item The fact that India does not feature among the top tourist destinations of the world may again have helped. [12] India's international tourist footprint is very small in comparison to France for example. The fact, however, remains that India features at rank 8 among the top 10 tourist destinations in Asia-Pacific.[12] The tourism is, however, localized---large parts of north and northeastern India (wherein impact of Covid-19 pandemic was lower) boast of very few international airports and fewer direct international flights. Significantly, Kerala, a state which continues to have a high number of cases has four international airports, while by comparison there is one in the state of Bihar and two in the state of Uttar Pradesh.[13] 
\end{enumerate}

\section{How India leveraged its strength}

Credit must go to the country's policy makers on their ability to identify its strengths and leverage these to their advantage.  
\begin{enumerate} 
\item \textbf{Lockdown and Zoning}: Use of non pharmaceutical interventions by imposing large-scale restrictions on gatherings where person-to-person transmission could occur was probably the most effective step taken by the Government. Some of these restrictions continue to be in place across large part of the country restricting social gatherings. As India's Prime minister said in one of his meetings with the Chief Ministers of different states at the end of April 2020, the government may move towards a ``smart lockdown''---with severe restrictions in affected districts, and partial lifting of restrictions in unaffected districts, along with the opening up of some sectors to meet the economic challenge. [14] This was done by demarcating the country into three types of zones---red, orange and green---depending on the scale of the Covid-19 outbreak. While no activity was allowed in the red zone, minimum activities like opening of limited public transport, harvesting of farm products were allowed in orange zones (where only few cases had been found in the recent past), and further relaxation like opening of MSME industries falling with in-house lodging facilities for employees with proper maintenance of social distance was allowed in green zones. [14] India is continuing to rely on this model of both restriction and regulation on the one hand and of relaxation on the other. The state of Maharashtra which has seen a rise in cases again is contemplating imposition of a new lockdown. [15] In a statement on 17th of February 2021, the chief minister of the state of Maharashtra minced no words while addressing a meeting with the Revenue Commissioners and Collectors through video conferencing, when he said that ``It is up to the people to decide whether they want the lockdown back or roam freely with some restrictions and if people do not wear masks or follow health rules, then the district and police administration has a responsibility to strictly enforce these rules. They must take strict punitive and necessary action without showing any leniency,''  Thus, imposing restriction and localizing cases to specific geographical locations is continuing to be seen as an effective tool to counter the impact of the pandemic.
\item \textbf{Restriction on travel}: For the large part of 2020, the country saw  restrictions on travel. Travel was regulated through issue of e-pass with border checks in place on state borders across the country. The regulation of travel continues into 2021 as travel has not been fully de-regulated. In a few states/union territories like Jammu \& Kashmir or Uttarakhand for example, every traveler is registered before entry or either needs to show Covid-19 negative test report or undergo a test for the infection. As late as 31st of January 2021, all 20 districts of the Union territory of J\&K were listed in orange zone with only regulated activities allowed.[16] The guidelines on international travel continue (as of 17th February 2021) to be in place with travelers requiring to undergo quarantine after entering India in addition to production of RT-PCR negative test report.
\item \textbf{Immediate isolation and contact tracing on testing positive}: The country continues to use its strategy of restricting mobility and isolating all individuals who test positive to Covd-19. Restrictions are imposed on their movement and contact tracing is conducted. All primary contacts identified as high-risk contacts (as per the defined protocol) are also quarantined and tested. The movements of contacts are restricted till they are confirmed to be negative. Quarantines of locations (not just of individuals) are also applied: The area/building or site is sealed off if the cases come in a cluster. In one such case when 14 inmates of an old age home in the state of Himachal Pradesh tested positive on 31st of January 2021, a 50 meter area around the old age home was completed sealed off and contained. [17] 
\item \textbf{School Closure}: For an extended period of time, all Indians up to age 25 were at home, there were no schools and no colleges open. Recent reopening is partial and occurs in a context of very low number of cases. The current reopening strategy, due to its recent nature, remains untested in its epidemiological consequences [21]    
\item \textbf{Industry response}: Another area wherein India showed tremendous progress within a short period of time was manufacturing. The industry responded to the needs of the pandemic like never before. From a country producing no high-end masks and personal protective equipment to its capability of being able to supply to the world, the shift was swift and decisive. [18] In fact, the production of personal protective equipment (PPEs) went from zero to the second-highest in the world within two months. The incentives by the Government worked as a boost as it itself went into procuring and distributing ventilators.
\item \textbf{Laboratory Testing Capacity}: From a single lab capable of doing RT-PCR for Covid-19 to more than 2300 (in both Government and Private sector) has been a significant public health initiative.[4] The addition of a greater number of labs has reduced the test result return time, strengthening the isolation/quarantine strategy.
\item \textbf{Awareness}: The public awareness drives with cautionary messages replacing ring tunes on phones to imposing fines for not wearing mask in public, public communication and response efforts moved in sync. More awareness meant that the general population willingly submitted to the restriction guidelines. Typical Indians travelled significantly less in 2020 than the past few years.
\item \textbf{High density urban areas}: The greatest challenge to pandemic response is the areas of urban deprivation. India's response therein involved local intensification and refinement of the lockdowns, travel restrictions, rapid case identification, and public communication found in other locations. For, e.g. Dharavi, one of the largest slums in Asia (2.1 km$^2$, about 1 mile on a side) and 1M people, the area is treated as a separate administrative zone with an Assistant municipal commissioner in charge. $3,60,600$ were screened, 9 municipal dispensaries and 350 private clinics were pressed into service, 225 community toilets, 100 public toilets, 125 MHADA toilets were disinfected daily. Anybody in these blocks that has any problems is reported to the 350 local health workers, they respond and provide isolation away from home. Quarantine facilities were established in schools, hostels etc. Communication is strong across the community that if there are problems to get tested and isolated as soon as symptoms are found. There is restricted entry and exit from these areas. Similarly, for multiple zones of slums in Delhi.[20]
\item \textbf{Vaccines}: Through what is being recognized as the largest vaccination drive across the world, the government of India on 16th of January started vaccinating its healthcare professionals, riding largely on two vaccines developed by Indian vaccine manufactures.[9] This action, reflecting an all measures approach, is expected to have a increasingly significant impact on the outbreak. 

\end{enumerate} 

\section{What India needs to continue doing}

In order to further reduce the number of daily cases of Covid-19 being reported, India needs to strengthen what it has been doing so far. It needs to continue with regulation/restriction in areas reporting cases while allowing return to normalcy in areas reporting no cases. It also needs to respond quickly to every single case of Covid-19 reported from any part of the country by isolating the case, carrying out contact tracing and placing restriction of movement on those suspected to carrying the disease. Also, it needs to strengthen its Covid-19 vaccination strategy as it does not have a general adult vaccination program in place. As the country intends to vaccinate 300 million individuals by July 2021, a vaccination strategy integrated into the pandemic response program may be helpful. 

Still, the difference between success and failure in pandemic response rests on decisiveness of action when numbers of cases are small. Recent numbers of new cases in India have leveled at around 11,000 cases per day with approximately half in Kerala. The outbreak there should be finally controlled. Elsewhere, and there, the urgency is great to complete the elimination process so that a new surge does not occur across India. Should India join countries in Asia and Oceania in elimination, $50\%$ of the world would become COVID free and an example for emulation can be achieved that is surely unexpected by western observers.

\section{Lessons for the world}

At the end of Spanish flu of 1918, the Sanitary Commissioner of India reported ``Transportation systems aided the spread of the disease as the railway played a prominent part as was inevitable''. [7] The Covid-19 pandemic has not behaved differently. Travel and time of local response are the key components of this pandemic's outcomes. The message to go out from India therefore is ``restrict travel in areas wherein cases are there, isolate cases and don't allow them to transmit infection''. The states that continue (to date) to have some restrictions on travel did better than others in India. Kerala, was the one state that opened up early, as early as April 2020, and distinct from other states continues to have significantly more cases. Probably also the fact that Kerala saw the largest number of expatriates coming back from outside the country could be a contributing factor. Of almost 8 lakh expatriates coming to India more than $25\%$ belonged to Kerala.  

\section{Conclusions} 

Travel and time could be the key for controlling pandemics. India, with less travel externally and internally than its western counterparts due to both intrinsic conditions and public health measures gave itself more time to get the infection under control than much of the western world. Under these conditions, where localized restrictions are not needed, relaxing restrictions can be done without preventing effective control measures in other locations, resulting in overall ongoing reduction in cases. That those who are not aware of the geographic dynamic of response efforts do not recognize it is not surprising, but the scientific and public health understanding should be clear.

\end{document}